\documentclass[11pt]{article}
\usepackage[french,english]{babel}
\usepackage{amsmath}
\usepackage{amsfonts}
\usepackage{amssymb}
\usepackage{times}
\begin{document}
%
%
March 27th 2012 \hfill
\vskip 2cm
{\baselineskip 20pt
\begin{center}
{\bf MIXING AND 1-LOOP FLAVOR STRUCTURE OF FERMIONIC CURRENTS

IN THE STANDARD MODEL OF ELECTROWEAK INTERACTIONS}

\end{center}
}
\vskip .2cm
\centerline{ B.~Machet
     \footnote[1]{LPTHE tour 13-14, 4\raise 3pt \hbox{\tiny \`eme} \'etage,
          UPMC Univ Paris 06, BP 126, 4 place Jussieu,
          F-75252 Paris Cedex 05 (France),\\
         Unit\'e Mixte de Recherche UMR 7589 (CNRS / UPMC Univ Paris 06)}
    \footnote[2]{machet@lpthe.jussieu.fr}
     }
\vskip 1cm

{\bf Abstract:} We show that, unlike mass matrices, the fermionic gauge currents
of the Standard Model exhibit, at the quantum level, remarkable $SU(2)_f$
 flavor properties at the observed values of the mixing angles. They
accommodate all measured mixing for three families of quarks, and, for
neutrinos, maximal $\theta_{23}$, quark-lepton complementarity
$\tan 2\theta_c=1/2 \leftrightarrow \tan 2\theta_{12}=2$, and a not so
small $\sin^2 2\theta_{13} = .267$
within the present $90\%\; c.l.$ interval of the $T2K$ experiment.

\smallskip


\section{Introduction}

We wish to combine two fairly old concepts which have
long been proven fruitful: Current Algebra \cite{CA} and Gell-Mann's
use of  ``strong $SU(3)$'' symmetry \cite{GM}):\newline
* much physics is conveyed by {\em currents} and is strongly constrained
 by their algebraic properties;\newline
* the laws of transformations of terms in a Lagrangian which {\em break}
a given symmetry carry important physical information;
this is how Gell-Mann deduced his famous mass formul{\ae} for hadrons.

The currents under concern here are the gauge currents of the Standard Model
of electroweak interactions, and the symmetry group flavor $SU(2)_f$ and
its $U(1)$ subgroup  of flavor rotations between pairs of fermions
carrying the same electric charge.

\section{The Standard Model in flavor space for two generations of
quarks}

\subsection{The classical Lagrangian}

The basics are more easily explained with two generations of quarks only,
$(u,d)$ and $(c,s)$.
These are the usual $SU(2)_L$ doublets, that we immediately reshuffle
into the two flavor doublets $(u_f,c_f)$ and $(d_f,s_f)$. We use the
subscript ``$_f$'' to denote (bare) flavor eigenstates. No transition
exists at this order between $c_f$ and $u_f$, nor between $s_f$ and $d_f$:
these states are (classically) orthogonal.

It is convenient
to embed the (global part of the) electroweak group into the chiral
flavor group $U(4)_L \times U(4)_R$. The electroweak $SU(2)_L$
generators write then at
the classical level ($1$ is the $2 \times 2$ unit matrix)
\begin{equation}
T^3 = \frac12 
\left(\begin{array}{ccc}
1   & \vline &   \cr  
\hline
& \vline & -1    \end{array}\right),
T^+ = \left(\begin{array}{ccc}
 &   \vline &  1 \cr 
\hline
& \vline & \end{array}\right),   
T^- = \left(\begin{array}{ccc}
 &   \vline   &  \cr 
\hline
1 & \vline &  \end{array}\right) .
\label{eq:TTTflav}
\end{equation}
Gauge currents form an $SU(2)_L$ triplet: they are obtained by sandwiching
the generators in eq.~(\ref{eq:TTTflav}) between $((u,c),(d,s))$ fermions and inserting
the appropriate $\gamma$ matrices.
Neutral currents are controlled by $T^3$; they form two singlets
of $SU(2)_f$: classically there is no flavor changing neutral current (FCNC)
(non-diagonal terms are vanishing)
nor any violation of universality (diagonal terms are two by two identical).

The standard Cabibbo
 phenomenology makes use of Yukawa couplings to induce two mass
matrices for the fermions, one for $(u_f,c_f)$, one for $(d_f,s_f)$. After
they are diagonalized, which introduces two mixing angles $\theta_u$ and
$\theta_d$ the $SU(2)_L$
generators write in the space of mass eigenstates $(u_m,c_m), (d_m,s_m)$
\begin{equation}
T^3 = \frac12 
\left(\begin{array}{ccc}
1   & \vline &   \cr  
\hline
& \vline & -1    \end{array}\right),
T^+ = \left(\begin{array}{ccc}
 &   \vline &  C \cr 
\hline
& \vline & \end{array}\right),   
T^- = \left(\begin{array}{ccc}
 &   \vline   &  \cr 
\hline
C^\dagger & \vline &  \end{array}\right) ,
\label{eq:TTTm}
\end{equation}
in which $C$ is the unitary Cabibbo matrix $C=\left(\begin{array}{rr}
\cos\theta_c & \sin\theta_c \cr -\sin\theta_c & \cos\theta_c
\end{array}\right)$,\quad $\theta_c=\theta_u-\theta_d$.

\subsection{Its symmetries}

Let ${\cal R}(\omega) = \left(\begin{array}{rr}
\cos\omega & \sin\omega \cr -\sin\omega & \cos\omega
\end{array}\right)$. The Cabibbo matrix $C$ and the fermion masses stay
 unchanged if one makes
arbitrary and independent flavor rotations
$\left(\begin{array}{cc}u_f \cr c_f \end{array}\right) \rightarrow
{\cal R}(\theta)\left(\begin{array}{cc}u_f \cr c_f \end{array}\right), 
\left(\begin{array}{cc}d_f \cr s_f \end{array}\right) \rightarrow
{\cal R}(\phi)\left(\begin{array}{cc}d_f \cr s_f \end{array}\right)$.

However, preserving $SU(2)_L$ symmetry requires that fermions belonging
to the same $SU(2)_L$ doublet be transformed with the same phase; thus
$\theta$ must be equal to $\phi$, which leaves a $U(1)$ invariance.

The Cabibbo angle $\theta_c$ stays arbitrary, also corresponding to an
arbitrary $U(1)$ rotation. This symmetry (arbitrariness) gets broken in
nature, but there is no hint at the classical level of how it is broken.

\subsection{Notations}

It is convenient for the following  to introduce the set of
three flavor $SU(2)_f$ generators, which depend on a (mixing) angle $\alpha$:

\begin{equation}
{\cal T}_x(\alpha)=\frac12
\left(\begin{array}{rr} \cos 2\alpha & \sin 2\alpha \cr
\sin 2\alpha & -\cos 2\alpha \end{array}\right),\quad
{\cal T}_y = \frac12 \left(\begin{array}{cc}
 & -i \cr i & \end{array}\right), \quad
{\cal T}_z(\alpha) = \frac12 \left(\begin{array}{rr}
\sin 2\alpha & -\cos 2\alpha \cr -\cos 2\alpha & -\sin 2\alpha
\end{array}\right) .
\label{eq:group}
\end{equation}

A 2-dimensional flavor rotation also writes ${\cal R}(\alpha) =
e^{2i\alpha {\cal T}_y}$ and ${\cal T}_{x,z}(\alpha)$ satisfy the relation

\begin{equation}
{\cal R}^\dagger(\omega)\, {\cal T}_{x,z}(\alpha)\,{\cal R}(\omega) =
{\cal T}_{x,z}(\alpha + \omega).
\label{eq:RTT}
\end{equation}

\subsection{At 1-loop} 

$s_m \rightarrow d_m, c_m \rightarrow u_m$ transitions spoil the
orthogonality of bare flavor states and Cabibbo
phenomenology \cite{DuMaVy1}, unless one introduces a series of counterterms
\cite{Shabalin}. They are
both kinetic and mass-like, with both chiralities.  In the $(d,s)$ sector,
they write (the expressions are analogous in the $(u,c)$
sector) \cite{DuMaVy2}
\begin{equation}
-A_d\; \bar d_m \, p\!\!/(1-\gamma^5)\, s_m - B_d\; \bar d_m
(1-\gamma^5) s_m
-E_d\; \bar d_m \, p\!\!/(1+\gamma^5)\, s_m - D_d\; \bar d_m (1+\gamma^5)
s_m,
\label{eq:AC}
\end{equation}
with

\vbox{
\begin{eqnarray}
A_d = \frac{m_d^2\, f_d(p^2 = m_d^2)- m_s^2\, f_d(p^2 = m_s^2)}{m_d^2 - m_s^2},&&
E_d = \frac{m_s m_d \left(f_d(p^2 = m_d^2) - f_d(p^2 = m_s^2)\right)}{m_d^2 -
m_s^2},\cr
&& \cr
B_d = -m_s\, E_d,&& D_d= -m_d\, E_d,
\label{eq:BD}
\end{eqnarray}
}

In the unitary gauge for the $W$ boson and dimensionally regularized
\begin{equation}
f_d=g^2 \sin\theta_c\cos\theta_c (m_c^2 - m_u^2)\int_0^1 dx \left[
\frac{2x(1-x)}{\Delta(p^2)} + \frac{p^2 x^3(1-x)}{M_W^2 \Delta(p^2)}
+\frac{x+3x^2}{M_W^2 \Delta(p^2)^{2-n/2}}\Gamma(2-n/2)\right].
\label{eq:f}
\end{equation}
$n=4-\epsilon$ is the dimension of space-time,  $\Delta(p^2) =
(1-x)M_W^2 + x\frac{m_u^2 + m_c^2}{2} -x(1-x)p^2$, and $\Gamma$ is the
Euler Gamma function which satisfies in particular
 $\Gamma(\epsilon/2) = 2/\epsilon -\gamma + \ldots$
where $\gamma \approx 0.5772\ldots$ is
the Euler constant. We shall come back later to more specific properties.

\subsection{1-loop mixing and currents}\label{subsec:1loop}

Adding counterterms to the bare Lagrangian modifies its diagonalization and
thus mixing matrices. We define the mixing matrix ${\cal C}_d$ in the
left-handed $(d,s)$ sector  by $\left(\begin{array}{c} d_f \cr s_f
\end{array}\right) = {\cal C}_d \left(\begin{array}{c} d_m \cr
s_m\end{array}\right)$, in which $d_m, s_m$ are the new mass eigenstates
(similar expressions hold in the $(u,c)$ sector).

Finding the new expressions of gauge currents is simple \cite{DuMaVy2} thanks to a
straightforward consequence of $SU(2)_L$ gauge invariance which dictates that the
derivatives in the kinetic-like counterterms  be replaced by the $SU(2)_L$
covariant derivatives. In a first step one shows that they are controlled
by the unit matrix in the new mass basis. Stepping back to the original
bare flavor basis, one finds \cite{DuMaVy2} that they are controlled by
 the $2\times 2$ matrices
$({\cal C}_d^{-1})^\dagger {\cal C}_d^{-1} = 1 + 2 A_d {\cal
T}_z(\theta_d)$ in the $(d,s)$ sector,
and $({\cal C}_u^{-1})^\dagger {\cal C}_u^{-1} = 1 + 2 A_u {\cal
T}_z(\theta_u)$ in the $(u,c)$ sector.
 These expressions exhibit in particular the non-unitarity
of the mixing matrices ${\cal C}_{u,d}$ connecting bare flavor states to 1-loop
mass states, which has already been extensively commented upon
\cite{DuMa1}\cite{DuMaVy2}\cite{MaNoVy}\cite{Novikov}.

Likewise, the charged current writes \cite{DuMaVy2}
$\left(\begin{array}{cc} \bar u_f & \bar c_f\end{array}\right)
\left[ 1 +A_u {\cal T}_z(\theta_{u}) +A_d
{\cal T}_z(\theta_{d}) \right]\gamma^\mu_L
\left(\begin{array}{c} d_f \cr s_f\end{array}\right)$, which is also
controlled, in the bare flavor basis, by a non-unitary matrix
\footnote{It is important to recall that the mixing matrices ${\mathfrak C}_d,
{\mathfrak C}_d$ and ${\mathfrak C}$ which connect 1-loop mass eigenstates
to 1-loop flavor states (instead of bare flavor states) are unitary and
satisfy the relation ${\mathfrak C} ={\mathfrak C}_u^\dagger{\mathfrak
C}_d$ \cite{DuMaVy2}. The standard Cabibbo (CKM) phenomenology is
indeed restored at 1-loop by the counterterms.}.

The back-reaction of introducing 1-loop counterterms is accordingly
 that the classical mixing angles $\theta_u$ and $\theta_d$ now occur
inside the gauge currents expressed in the bare flavor basis. Some formal
structure starts to appear through the $SU(2)_f$ generator
${\cal T}_z$. It is also important that only the $A$
counterterms occur, because of the renormalization freedom attached to
them, that we shall soon focus on.

\subsection{Flavor rotations}

Like in the classical case, we perform flavor rotations, respectively
with angles $\theta$ on
$(u_f,c_f)$ and $\phi$ on $(d_f,s_f)$. Since the classical Cabibbo angle
$\theta_c$ is unchanged, one easily shows that the counterterms also stay
identical. The neat effect of these arbitrary transformations is finally
the one obtained  naively by transforming the neutral and charged currents
given above and using the relations (\ref{eq:RTT}). The neutral currents get now
controlled in bare flavor space by $1 + 2 A_d {\cal
T}_z(\theta_d + \phi)$ in the $(d,s)$ sector,
and $ 1 + 2 A_u {\cal T}_z(\theta_u + \theta)$ in the $(u,c)$ sector. The
charged current becomes (omitting the fermions and the $\gamma$ matrix) 
$ 1 +A_u {\cal T}_z(\theta_u +(1/2)(\theta+\phi)) +A_d
{\cal T}_z(\theta_d +(1/2)(\theta+\phi))$. Like before, going through all
the steps, one can show that the Cabibbo angle stays unchanged, whatever
$\theta$ and $\phi$, and so do the mass eigenvalues. So, ``physics is
unchanged''.
 Arguing with $SU(2)_L$ symmetry to
impose $\phi=\theta$, the arguments of all ${\cal T}_z$
generators become $\theta_{u,d} + \phi$. 
The arbitrariness of $\phi$ reflects an up-to-now unbroken $U(1)$ invariance.

At 1-loop, neutral currents are no
longer flavor singlets of $SU(2)_f$: there exist, in bare flavor
space, both FCNC's and violations of universality (differences among
diagonal terms); {\em flavor rotations continuously transform FCNC's into
violation of universality and  vice-versa}.

\subsection{Left-over freedom}

Since an arbitrary rotation by $\phi$ does not change ``physics'', we can
choose $\phi=-\theta_u$, which aligns classical flavor and mass $(u,c)$
fermions.

The  counterterms $A_d, B_d, D_d, E_d$ satisfy the
conditions $B_d = cst$ (thus $E_d = cst$ and $E_d=cst$ and there is no
freedom on $B,E,D$). $A_u$ and $A_d$ have both a pole in $n-4=\epsilon$ but
the combination $(m_c^2 - m_u^2)A_u - (m_s^2-m_d^2)A_d$ is finite
\footnote
{It is $g^2 \sin^2\theta_c\frac{(m_c^2-m_u^2)(m_s^2-m_d^2)}{m_W^4}\left[
\frac{27}{12}(m_c^2+m_u^2) + \frac{23}{24}(m_s^2+m_d^2)\right]\left(1+{\cal
O}(m^2/m_W^2)\right)$.}.
This shows in particular that at least one non unitary mixing matrices occur
in one of the two channels $(u,c)$ and $(d,s)$.
One has the freedom to renormalize to
$0$ one among the two counterterms $A_u$ or $A_d$. Let us choose $A_u=0$;
the alignment of flavor and mass states that we imposed classically by
choosing $\phi=-\theta_u$ is maintained at 1-loop. Then,
$A_d$ can only be non-vanishing $A_d \approx -g^2\sin^2\theta_c
\frac{27}{12}\left(\frac{m_c}{m_W}\right)^4 \approx 1.24\, 10^{-7}$.

After these manipulations, the neutral currents are respectively controlled by
$1$ in the $(u,c)$ channel, $1+2A_d{\cal T}_z(\theta_d-\theta_u)$ in the
$(d,s)$ channel, and the charged currents by $1+A_d{\cal
T}_z(\theta_d-\theta_u)$, such that
they all only depend on the arbitrary Cabibbo angle
 through the ${\cal T}_z(\theta_c)$ generator.

\subsection{How flavor rotations get broken}

Once all freedom has been used, the arbitrariness of $\phi$ has become
that of the Cabibbo angle.

We noticed that classical neutral gauge currents only project on $SU(2)$ flavor
singlets. The simplest extension is that quantum corrections add to it a
part  proportional to a triplet of $SU(2)_f$, that is\newline
 $\left(\begin{array}{rr} \bar d & \bar s\end{array}\right)
\left[
 \left(\begin{array}{rr}0 & \pm 1 \cr 0 & 0 \end{array}\right),
\left(\begin{array}{rr} 0 & 0 \cr \pm 1 & 0 \end{array}\right),
\frac12\left(\begin{array}{rr} 1 & 0 \cr 0 & -1\end{array}\right)
\right] \gamma^\mu_L\left(\begin{array}{c}d \cr
s\end{array}\right)$,\newline
which corresponds in the Lagrangian to 
$\left(\begin{array}{cc} \bar d & \bar
s\end{array}\right) \left(\begin{array}{cc} 1/2 & \pm 1 \cr \pm 1 &
-1/2\end{array}\right)\gamma^\mu_L
\left(\begin{array}{c}d \cr s\end{array}\right)$ (the three components of
the triplet are coupled to the same gauge boson, unlike what happens for
the $SU(2)_L$ triplet of electroweak currents (\ref{eq:TTTflav})).

Comparing with the expressions obtained in subsection \ref{subsec:1loop}
and matching the term proportional to ${\cal T}_z$ 
\footnote{The general expression of ${\cal T}_z$ is given in
eq.~(\ref{eq:group}).}
 with the formul{\ae}
above, one gets the relation
$\tan 2\theta_c = \pm 1/2$, which corresponds with very good accuracy to the
measured value of the Cabibbo angle.
The choices that we made in order to align at 1-loop $(u,c)$ flavor 
and mass states
ensure that the same angle that occurs in the $(d,s)$ neutral current also
 occurs inside charged currents, which get similar structures
\footnote{The $U(1)$ breaking occurs in such a way that the violation of
universality $1/2 - (-1/2)=1$ is identical to the ``amount'' of FCNC which
is also $1$. So, flavor rotations, which, as we stressed, continuously transform
 the former into the latter, are broken in such a way that the two violations
occur with the same strength.}.

\subsection{The last generator: $\boldsymbol{{\cal T}_x}$ and the mass matrix}

Of the three $SU(2)_f$ generators, we have up to now only used ${\cal
T}_y$, which, by exponentiation, triggers flavor rotations, and ${\cal
T}_z$ which controls gauge currents. It is natural to look for ${\cal T}_x$
in the mass terms since it is the third element of the process: diagonalization
$\rightarrow$ mixing matrix $\rightarrow$ gauge currents. And, indeed,
supposing, to simplify, that the mass matrix is symmetric, it
writes\newline
$M=\left(\begin{array}{cc} a & c \cr c & b\end{array}\right) =
\frac{m_2+m_1}{2} +  (m_2-m_1) {\cal T}_x(\theta)$, where $m_1$
and $m_2$ are the eigenvalues of $M$ and $\tan 2\theta= \frac{2c}{a-b}$.
This corresponds
 to the decomposition of the symmetric non-degenerate matrix $M$:
$M= m_1 P_1 + m_2 P_2$, where $P_1 = \frac12(1 - 2{\cal T}_x(\theta))$ and
$P_2 = \frac12(1+2{\cal T}_x(\theta))$ are the projectors on the sub-spaces
of  eigenvectors of $M$.

 So, the whole set \{masses, currents\}
of the Standard Model is controlled by the $SU(2)_f$ flavor group with
generators given by (\ref{eq:group}).

Most efforts have been devoted to establishing connections between
the mass spectrum $(m_1,m_2)$ and the Cabibbo angle (see for example
\cite{Giunti}).
This is probably not the appropriate way to proceed for at least two
reasons:\newline
* any homographic transformation on a mass matrix $M: M\rightarrow
\frac{\alpha M + \beta}{\delta M + \gamma}$ leaves unchanged the
eigenvectors of $M$ and thus the mixing angles: an infinite number of
different mass matrices, with different eigenvalues, therefore correspond to the
same mixing angle;\newline
* the structure that we found within gauge currents stays
hidden at the level of the mass matrix. Indeed, the $(d,s)$ mass matrix
corresponding to $\tan 2\theta_d = \frac12$  is
$M = \frac{m_s+m_d}{2} + \frac{m_s-m_d}{\sqrt{5}}\left(\begin{array}{rr}
1 & \frac12 \cr \frac12 & -1\end{array}\right) = \left(\begin{array}{cc}
m_s(\frac12+\frac{1}{\sqrt{5}}) + m_d(\frac12-\frac{1}{\sqrt{5}}) &
\frac{m_s-m_d}{2 \sqrt{5}}\cr \frac{m_s-m_d}{2 \sqrt{5}} &
m_s(\frac12 -\frac{1}{\sqrt{5}})+ m_d(\frac12 +\frac{1}{\sqrt{5}})
 \end{array}\right)$. Though its part proportional to $(m_s-m_d)$ is
obtained from
the triplet part of neutral currents by a rotation $\theta \rightarrow
\theta+\frac{\pi}{4}$, it has as a whole no conspicuous property.

\section{Three generations of quarks}

The simplest generalization of what has been done for two generations is to
consider, inside each flavor triplet of quarks, $(u,c,t)$ and $(d,s,b)$,
the three $2\times 2$ flavor rotations associated with the three quark pairs,
and the corresponding three sets of four currents,
for example [$\bar d \gamma^\mu_L s, \bar s
\gamma^\mu_L d, \bar s\gamma^\mu_L s, \bar d \gamma^\mu_L d$],
[$\bar d \gamma^\mu_L b, \bar b \gamma^\mu_L d, \bar b\gamma^\mu_L b,
 \bar d \gamma^\mu_L d$] and
[$\bar b \gamma^\mu_L s, \bar s
\gamma^\mu_L b, \bar s\gamma^\mu_L s, \bar b \gamma^\mu_L b$].
One then requests that each set decomposes
into a trivial singlet part + a triplet of the corresponding $SU(2)_f$.  
This  is akin to requesting that the matrix $({\cal C}_d^{-1})^\dagger
{\cal C}_d^{-1}$ which controls neutral currents, be of the form
\footnote{ Eq.~(\ref{eq:3gen}) also writes
$({\cal C}_d^{-1})^\dagger {\cal C}_d^{-1}= \newline
1 +
(\alpha-\beta)\left(\begin{array}{rrr} 1/2 & \pm 1 & 0 \cr
 \pm 1 & -1/2 & 0 \cr 0 & 0 & 0 \end{array}\right)
+(\beta-\gamma)\left(\begin{array}{rrr} 0 & 0 & 0 \cr
0 & 1/2 & \pm 1 \cr 0 & \pm 1 & -1/2  \end{array}\right)
+(\alpha-\gamma)\left(\begin{array}{rrr} 1/2 & 0 & \pm 1 \cr
0 & 0 & 0 \cr \pm 1 & 0 & -1/2 \end{array}\right) \newline
+\frac{\alpha}{2}\left(\begin{array}{rrr} 0 & & \cr & 1 & \cr & & 1
 \end{array}\right)
+\frac{\beta}{2}\left(\begin{array}{rrr} 1 & & \cr & 0 & \cr & & 1
 \end{array}\right)
+\frac{\gamma}{2}\left(\begin{array}{rrr} 1 & & \cr & 1 & \cr & & 0
 \end{array}\right)$, making explicit its decomposition on $SU(2)_f$ singlets 
and triplets.}
\cite{DuMaVy2}

\begin{equation}
({\cal C}_d^{-1})^\dagger {\cal C}_d^{-1}=
1 + \left(\begin{array}{ccc}
  \alpha  & \pm(\alpha -\beta) & \pm(\alpha -\gamma)\cr
  \pm(\alpha -\beta) & \beta & \pm(\beta-\gamma)    \cr
  \pm(\alpha - \gamma) & \pm(\beta-\gamma) & \gamma
  \end{array}\right).
\label{eq:3gen}
\end{equation}

Decomposing the $3\times 3$ mixing matrix ${\cal C}_d$ as a product of three
``quasi-rotations'' acting in each of the three 2-dimensional flavor
sub-spaces, and introducing, like in the $2\times 2$ case, two mixing angles
for each of them, this yields a set of trigonometric equations which
have been investigated and solved in \cite{DuMaVy2}. Even if the whole set
of solutions was not produced, it was shown that it includes values of the
mixing angles  in remarkable  agreement with observation.

This accordingly supports the conjecture that $2\times 2$ flavor rotations
get spontaneously broken such that  the corresponding neutral currents
decompose onto a singlet + a (small) triplet component of $SU(2)_f$.
At this point it is useful to note that the decomposition (\ref{eq:3gen})
is more restrictive than a singlet + octet of $SU(3)_f$, because the
Gell-Mann matrices $\lambda_2, \lambda_5$ and $\lambda_7$ do not appear.

It is still more hopeless than for two generations
 to look for any particular structure in the mass matrix.

\section{Neutrinos}

\subsection{Where do they take their maximal mixing from?}

$\bullet$\ The perturbative vacuum of the Standard Model is unstable
while the true vacuum, away from which one can safely quantize small
variations,  cannot be reached perturbatively from the former. 
Let us transpose this type of consideration to mixing angles.
The usual ``classical solution'' corresponds here to a
unitary ${\cal C}_d = {\cal R}(\theta_d)$ mixing matrix with
an arbitrary $\theta_d$.  We call it "Cabibbo-like''. 
Can there be other classical solutions? If we parametrize the mixing matrix
by a rotation ${\cal R}(\theta_d)$, we have no chance whatsoever to
eventually find them. We should instead parametrize it in a general non-unitary
form, since we know that this is how it is ultimately realized, then find
all possible solutions at the ``classical limit''
 ${\cal C}_d^\dagger {\cal C}_d\rightarrow 1$.
This has been done in \cite{DuMa1} where we parametrized (for two
generations) \newline
${\cal C}_d= \left(\begin{array}{rr}
e^{i\alpha}c_1 & e^{i\delta}s_1\cr -e^{i\beta}s_2 & e^{i\gamma}c_2
\end{array}\right),\quad c_1=\cos\theta_1, s_2=\sin\theta_2\  etc$.
The first type of solutions are ``Cabibbo-like'' and correspond to\newline
*\ $\theta_2 = \theta_1 + k\pi$ associated with
$e^{i(\alpha-\delta)}=e^{i(\beta-\gamma)}$,\newline
 or to \newline
*\ $\theta_2 = -\theta_1 + k\pi$ associated with
$e^{i(\alpha-\delta)}=-e^{i(\beta-\gamma)}$; 

the second type of solutions are sets of discrete values
\footnote{which still satisfy $\theta_2 = \pm \theta_1 + k\pi$ and are thus
superposed to the continuous sets of Cabibbo-like solutions.}\newline
*\ [$\theta_1=(k-n)\frac{\pi}{2}, \theta_2=(k+n)\frac{\pi}{2}$] or
[$\theta_1=\frac{\pi}{4}+(n-k)\frac{\pi}{2},
\theta_2=\frac{\pi}{4}+(n+k)\frac{\pi}{2}]$, both associated with
$e^{i(\alpha-\delta)}=e^{i(\beta-\gamma)}$,\newline
*\ [$\theta_1=(n-k)\frac{\pi}{2}, \theta_2=(n+k)\frac{\pi}{2}$] or
[$\theta_1=-\frac{\pi}{4}+(k-n)\frac{\pi}{2},
\theta_2=\frac{\pi}{4}+(k+n)\frac{\pi}{2}$], both associated with
$e^{i(\alpha-\delta)}=-e^{i(\beta-\gamma)}$, \newline
which include in particular the maximal mixing $\pm\frac{\pi}{4}$.

$\bullet$\ In the case of three generations,
the measured values of the neutrino mixing angles point at
two ``Cabibbo-like'' classical mixing angles
$\theta_{12}, \theta_{13}$, combined with a maximal $\theta_{23}$.
It is again simple to see how maximal mixing springs out there.
Let us do this in the case where the third mixing
angle $\theta_{13}$ is, for the sake of simplicity, taken to vanish.
One can even take a parametrization of the mixing matrix without the phases
that were introduced in the case of two generations \cite{DuMaVy2}:
\newline
${\cal C}^{-1}=
\left(\begin{array}{rrr}
1 & 0 & 0 \cr 0 & c_{23} & s_{23}\cr 0 & -\tilde s_{23} & \tilde c_{23}
\end{array}\right)
\left(\begin{array}{rrr}
c_{12} & s_{12} & 0\cr -\tilde s_{12} & \tilde c_{12} & 0 \cr 0 & 0 & 1
\end{array}\right)$. It gives
the following expression for the symmetric matrix
 $({\cal C}^{-1})^\dagger {\cal C}^{-1}$ which controls neutral
currents:\newline
$ ({\cal C}^{-1})^\dagger {\cal C}^{-1} =\left(\begin{array}{ccc}
c^2_{12}+\tilde s^2_{12}(c^2_{23}+\tilde s^2_{23}) &
c_{12} s_{12} -\tilde c_{12} \tilde s_{12}(c^2_{23}+\tilde s^2_{23}) &
-\tilde s_{12}(c_{23}s_{23}-\tilde c_{23}\tilde s_{23}) \cr
c^2_{12}+\tilde s^2_{12}(c^2_{23}+\tilde s^2_{23}) &
s^2_{12} + \tilde c^2_{12}(c^2_{23} + \tilde s^2_{23}) &
-\tilde c_{12}(c_{23}s_{23}-\tilde c_{23}\tilde s_{23}) \cr
-\tilde s_{12}(c_{23}s_{23}-\tilde c_{23}\tilde s_{23}) &
-\tilde c_{12}(c_{23}s_{23}-\tilde c_{23}\tilde s_{23}) &
s^2_{23} + \tilde c^2_{23}
\end{array}\right)$,\newline
 where we used the abbreviated notations $\tilde s_{12}
= \sin\tilde\theta_{12}$ etc . The ``classical equations'', which correspond to
unitarity, are now five. Three constrain the non-diagonal elements to
vanish and two express the identity of diagonal elements.
Excluding $\tilde\theta_{12}=0$, the vanishing of the $[1,3]$ and $[2,3]$
entries of  $({\cal C}^{-1})^\dagger {\cal C}^{-1}$ requires $\sin
2\theta_{23} = \sin 2\tilde\theta_{23}$, that is: either $\tilde\theta_{23}=
\theta_{23} + k\pi$ or $\tilde\theta_{23}= \frac{\pi}{2}-\theta_{23} +
k\pi$. Choosing the second possibility yields, for example
$\tilde s_{23} = c_{23}$. Then, that the $[1,2]$ entry vanishes requires
$c_{12} s_{12} = 2 c^2_{23} \tilde s_{12}\tilde c_{12}$.
That the entries $[1,1]$ and $[2,2]$ are identical writes
 $c^2_{12} - s^2_{12} =
2c^2_{23}(\tilde c^2_{12}-\tilde s^2_{12})$, and, last, that 
$[2,2]$ = $[3,3]$
 requires $s^2_{12} + 2 c^2_{23}\tilde c^2_{12} - 2s^2_{23}$. These
last three equations let us note (a), (b) and (c).  Making the ratio
(a)/(b) yields $\tan 2\theta_{12} = \tan 2\tilde\theta_{12}$, which
requires $\tilde\theta_{12} = \theta_{12} + \frac{k\pi}{2} + n\pi$. The
same equations then yield $2c^2_{23}=1$, that is, $\theta_{23}$ {\em
maximal}. It
remains to verify that all equations are satisfied. They are indeed with
$\tilde\theta_{23} = \frac{\pi}{2}-\theta_{23} + k\pi$ and
$\tilde\theta_{12} = \theta_{12} + p\pi$. So, the classical solution
corresponds here to a Cabibbo-like $\theta_{12}$, to maximal $\theta_{23}$,
and to a (Cabibbo-like) $\theta_{13}=0$ (which is our simplifying
hypothesis).
True values of the mixing angles are then sought for as  solutions of
 eq.~(\ref{eq:3gen}) restricted to lie in the vicinity of this set.

\subsection{Mixing angles}

Solving the corresponding system of trigonometric equations in full
generality is quite an involved task but it is not our goal. 
We can instead keep the value of $\theta_{23}$ maximal  as it
has been obtained in the ``classical'' (unitary) limit, and solve in this
limit the equations for $\theta_{12}$ and $\theta_{13}$ given
by eq.~(\ref{eq:3gen}). 
It is also  convenient to start with $\theta_{13}=0$. With these two
constraints, one obtains \cite{DuMa2}\cite{DuMaVy2} for $\theta_{12}$
 the ``golden ratio value'' \cite{DuMa2}\cite{Strumia}
$\tan 2\theta_{12}=2$, which, associated with $\tan 2\theta_c=1/2$
 ensures the so-called ``quarks-leptons complementarity'' \cite{Raidal}.
Imposing then this value for $\theta_{12}$ and still keeping $\theta_{23}$
maximal, one solves the equations for $\theta_{13}$ coming from
eq.~(\ref{eq:3gen}) in the approximation that it is small
($\sin^2\theta_{13}<.1$). This gives \cite{DuMaVy2}
two possible values $\sin^2 2\theta_{13}=1.3\,10^{-4}$ and $\sin^2
2\theta_{13} =.267$. They lie within the recent $90\%\; c.l.$
 interval of the $T2K$ experiment $\sin^2 2\theta_{13} < .5$ \cite{T2K}.
In this framework accordingly perfectly accommodate
a value of $\theta_{13}$ not so small as it had
long been expected.

\section{Conclusion}

We have used  1-loop counterterms to motivate
explicit calculations of mixing angles at this order.
However the results that we have obtained are general and only rely on
two properties:\newline
* Quantum corrections always induce non-unitarity for the mixing matrices
connecting classical flavor eigenstates to 1-loop mass eigenstates
of non-degenerate coupled fermions
\cite{DuMa1} \cite{DuMaVy2} \cite{Novikov}; \newline
* Flavor rotations get broken in such a way that quantum corrections induce
$SU(2)_f$ triplets in 1-loop neutral currents
(expressed in terms of classical flavor eigenstates)
while they only involve singlets at the classical level.

Among open issues one can mention:\newline
* Why do, at the classical (unitary) limit, quarks 
exhibit three continuous ``Cabibbo-like'' mixing
angles  while neutrinos get a discrete value for one of them? We have seen
that, if one pays enough attention, all these choices are perfectly
legitimate (and also $ \pm\frac{\pi}{2}$ values), but we have
no dynamical explanation for them.
In particular, why does, seemingly, only $\theta_{23}$ 
get a finite ``classical value'' ? \newline
* The system of trigonometric
equations resulting from the symmetry constraints that we impose has many
solutions. That this set includes the subset of observed values gives no
hint concerning why this subset is chosen by nature and why other allowed
solutions are not.\newline
* Why does $SU(2)$ seems to play again an important role?
Is it more than an ``opportunistic'' (low-energy, at the best) symmetry only
introduced because it is the simplest that comes to mind? 
Should it be given a more fundamental status like that of a
(spontaneously broken or not) gauge symmetry?

Both the non-unitarity of mixing matrices in bare flavor space
 and the generation of $SU(2)_f$ triplets in neutral currents, which seemingly
control mixing angles, originate from  the existence of mass
splittings. However, there is no direct connection between masses and
mixing angles: the gap stays wide open and the mass spectrum remains apart.
Gell-Mann's achievement to find relations among  hadron masses from a specific
breaking pattern of $SU(3)$ $(u,d,s)$ flavor symmetry still waits for its
equivalent for fundamental fermions.
There lies most probably the realm of physics beyond the standard model.

\bigskip
\underline{\em Acknowledgments:} these notes outline the main ideas contained
in the works \cite{DuMaVy1}, \cite{DuMaVy2}, \cite{DuMa1} and \cite{DuMa2}
 done in collaboration with Q.~Duret and M.I.~Vysotsky, 
to which I address my warmest thanks.

\begin{em}

\end{em}

\end{document}